\documentclass[prb,twocolumn,floatfix]{revtex4}

\usepackage[dvips]{graphicx}
\usepackage{epsfig}

\newcommand{\half}{\mbox{$\frac{1}{2}$}}
\newcommand{\nod}{\noindent}
\newcommand{\bfa}[1]{\mathbf{#1}} 

\usepackage{subfigure}

\usepackage{amsmath}

\usepackage{natbib}

\begin{document}

\title{Fracture of Notched Single Crystal Silicon}

\date{\today}



\author{Nicholas P. Bailey}
\email{nbailey@fysik.dtu.dk}
\author{James P. Sethna}
\affiliation{Department of Physics, Cornell University, Ithaca, NY 14853}

\begin{abstract}
We study atomistically the fracture of single crystal silicon at atomically
 sharp notches with opening angles of 0 degrees (a crack), 70.53 degrees, 90 
degrees and 125.3 degrees.  Such notches occur in silicon that has been formed
 by etching
 into microelectromechanical structures and tend to be the initiation sites for
 failure by fracture of these structures. Analogous to the stress intensity
 factor of traditional linear elastic fracture mechanics which characterizes
 the  stress state in the limiting case of a crack, there exists a similar
 parameter $K$ for the case of the notch. In the case of silicon, a brittle
 material, this characterization appears to be particularly valid. We use 
three interatomic potentials: a modified Stillinger-Weber potential, the
 Environment-Dependent Interatomic Potential (EDIP), and the modified embedded 
atom method (MEAM). Of these, MEAM gives critical $K$-values
closest to experiment. In particular the EDIP potential leads to unphysical
ductile failure in most geometries. Because the units of $K$ depend on the 
notch angle, the shape of the $K$ versus angle plot depends on the units used.
In particular when an atomic length unit is used the plot is almost flat, 
showing---in principle from macroscopic observations alone---the association
of an atomic length scale to the fracture process.

\end{abstract}

\maketitle

\section{\label{introductionSilicon}Introduction}

Recently there has been experimental\cite{Dunn/others:1997,
Suwito/Dunn/Cunningham:1998, Suwito/others:1999} and 
theoretical\cite{Zhang:2002} interest in fracture in sharply
 notched single crystal silicon samples. Such samples have technological
 importance because silicon is a commonly used material in the fabrication of
 MEMS devices; the etching process used tends to create atomically sharp
 corners due to highly anisotropic etching rates.\cite{Suwito/others:1999} 
Failure in such devices is often a result of fracture which initiated at sharp 
corners.\cite{Suwito/Dunn/Cunningham:1997} In the case of a
 notch, there
 exists a parameter $K$ analogous to the stress intensity factor of traditional
 fracture mechanics, which parameterizes the elastic fields in the vicinity of
 the notch. Suwito et al.\cite{Suwito/Dunn/Cunningham:1998, Suwito/others:1999} have
 carried out a series of experiments which have (i)
 established the validity of the stress intensity factor as a fracture 
criterion in notched specimens and (ii) measured the
 critical stress intensities for several notch geometries. On the theoretical
 side Zhang\cite{Zhang:2002} has carried out an analysis which models the
 separation
 of cleavage planes by a simple cohesive law, and thereby derived a formula 
for the critical stress intensity as a function of notch opening angle. The
 material properties which enter this formula are the elastic constants and the
 parameters of the cohesive law, the peak stress $\hat\sigma$ and the work of
 separation $\Gamma_0$. This recent activity has prompted us to investigate
 the phenomenon of fracture in notched silicon using atomistic simulations: In 
this paper we present direct measurements of the critical stress
 intensity for different geometries (i.e., notch opening angles) and
compare them to the experimental results of Suwito et al. We apply a load by
 specifying a pure $K$-field of a given
strength (stress intensity factor) on the boundary of the system. In doing
 this we are effectively using the result of Suwito et al. that the notch 
stress intensity factor is indeed the quantity which determines fracture 
initiation, so we can ignore higher order terms in the local stress field.

\subsection{Elastic fields near a notch}

\begin{figure} 
\begin{center}
\epsfig{file=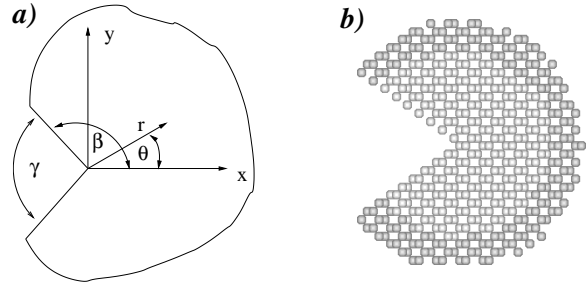,width=3.0 in}
\caption{\label{notchschematic}(a) Notch schematic and notation; (b) silicon 
crystal with a notch; darker layer is fixed boundary atoms. }
\end{center}
\end{figure}

The essential geometry of a notch is shown in Fig.~\ref{notchschematic}. The
notch opening angle is denoted $\gamma$ and the half-angle within the material,
which is the polar angle describing the top flank, is $\beta$ (thus 
$\beta=\pi - \gamma/2$). As discussed in detail by Suwito et 
al.,\cite{Suwito/Dunn/Cunningham:1998,Suwito/others:1999} it is fairly
 straightforward to solve the equations of anisotropic linear elasticity for a
 notched specimen. The formalism used is known as the Stroh 
formalism,\cite{Ting:1996} which is useful for dealing with materials with 
arbitrary
 anisotropy in arbitrary orientations, as long as none of the fields depend on
 the $z$ coordinate (this will be the out-of-plane coordinate; note that
 this does not restrict the deformation itself to be in-plane). Here we only
 consider mode I (symmetric) loading. The displacement and stress fields 
for a notch can be written as

\begin{equation}\label{analyticdisplacements}
u_i = K r^\lambda g_i(\theta)
\end{equation}

\begin{equation}\label{analyticstress}
\sigma_{ij} = K r^{\lambda-1} f_{ij}(\theta)
\end{equation}

\nod where $\lambda$ plays a role like an eigenvalue; its value is determined 
by applying the traction-free boundary conditions to the notch flanks. There is
an infinity of possible values for $\lambda$ of which we are interested in 
those in the range $0 < \lambda < 1$, which give rise to a singular stress
 field, often known as the $K$-field, at the
 notch tip. This is entirely analogous to the singular field near a crack tip,
 which is simply the limiting case where $\gamma$ goes to zero 
($\beta \longrightarrow \pi$), and $\lambda$ becomes one half. Further
 details of the Stroh formalism, as applied to the notch geometry, are given 
in appendix \ref{strohappendix}. The complete elastic solution involves the 
whole infinity of values for $\lambda$,
 corresponding to different multipoles of the elastic field. Negative values of
$\lambda$ correspond to more singular fields which are associated with 
properties of the core region stemming from the non-linear atomistic nature of 
this region; they do not couple to the far-field loading. $\lambda>1$ corresponds
 to fields which are less singular, and do not influence conditions near the notch-tip,
 since the displacements and stress vanish there. They are, however, essential to
 represent the full elastic field throughout the body, and ensure that boundary
 loads and displacements (whatever they may be) are correctly taken into 
account. This is the basis for asserting that only the $K$-field is important.
This field is unique among the multipoles in that it both couples to the far-field
loading and is singular at the notch tip. Thus the stress intensity factor must
characterize conditions at the crack tip, and therefore a critical value, $K_c$,
is associated with the initiation of fracture. The validity of this approach hinges
on the validity of linear elasticity to well within the region in which the $K$-field
dominates.

\begin{figure}
\begin{center}
\epsfig{file=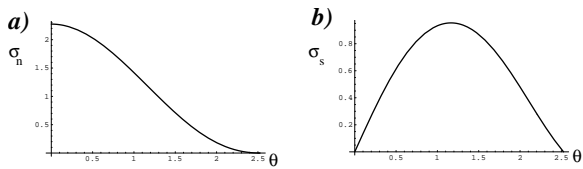,width=3.0 in}
\end{center}
\caption{\label{normalShearAngular}Normal (a) and shear (b) stresses on radial
 planes as functions of plane angle, for $\gamma=70^\circ$.}
\end{figure}

From Eq.~(\ref{analyticstress}) we see that the units of $K$ and therefore $K_c$ are
 $stress/length^{\lambda-1}$ which depends continuously on the notch angle $\gamma$
 through $\lambda$. Hence the shape of a plot of $K_c$ against notch 
angle depends on the units used to make the plot. In metric/SI units $K_c$
 changes by an order of magnitude between $70^\circ$ and $125^\circ$ whereas if
 an atomic scale unit of length is used the plot is nearly flat 
(Fig.~\ref{KvsAnglePlots}). The most interesting feature of this  
 is that it seems to provide a direct link
 from macroscopic measurements to a microscopic length scale. From a continuum 
point of view, one incorporates atomistic effects into fracture via a 
\textit{cohesive zone}, a region ahead of the crack tip where material cleaves
 according to a specified force-separation law. One of the parameters of such
 laws is the length scale---the distance two surfaces must separate before the
 attractive force goes to zero---which for a brittle material is an atomic length
 scale. It is this scale that one would identify from the plot of $K_c$ versus angle.
 Note that one can only identify a \textit{scale}, and not an actual length
 parameter, in particular because the different geometries that are involved in
 the plot involve different fracture surfaces, with presumably different
 force-separation parameters.

\begin{table}
\caption{\label{surfaceenergytable} Surface energies for silicon according to
 mSW, EDIP and MEAM potentials.}
\begin{center}
\begin{tabular}{|l|c|l|l|}
\hline
potential & surface & atomic units & SI units \\
\hline
mSW & 111 & 0.0906 & 1.3593 \\
mSW & 110 & 0.1110 & 1.6649 \\
mSW & 100 & 0.1570 & 2.3545 \\
EDIP & 111 & 0.06538 & 1.0475 \\
EDIP & 110 &  0.08194 & 1.3128\\
EDIP & 100 & 0.1320 & 2.1150 \\
MEAM & 111 & 0.07668 & 1.2285 \\
MEAM & 110 & 0.09030  & 1.4469 \\
MEAM & 100 & 0.08126 &  1.3019 \\
\hline
\end{tabular}
\end{center}
\end{table}

Fig.~\ref{normalShearAngular} shows the normal and shear stresses on radial
planes (perpendicular to the plane of the sample) emanating from the notch tip,
 for unit $K$ and $r$ (i.e., they are derived from the tensor $f_{ij}$ 
appearing in Eq.~(\ref{analyticstress})). The figures show the functions for
the $\gamma=70^\circ$ case; the other geometries have the same qualitative behavior.
Both stresses vanish at the maximum angles, corresponding to the notch flanks;
this is in accordance with traction-free boundary conditions.
What is most important to note is that the normal stress, which presumably is 
most relevant for cleavage on a radial plane, has its maximum at $\theta=0$.
The shear stress, which is relevant for possible slip behavior (dislocations)
 which could compete with cleavage as a means of relieving stress, is zero at
$\theta=0$, and has a maximum at intermediate angles. If there is an easy crystal slip
plane in the vicinity of the maximum, slip could conceivably compete with
cleavage.

\section{Simulation}

\begin{figure}
\begin{center}
\subfigure[$K=0.3064$]{\epsfig{file=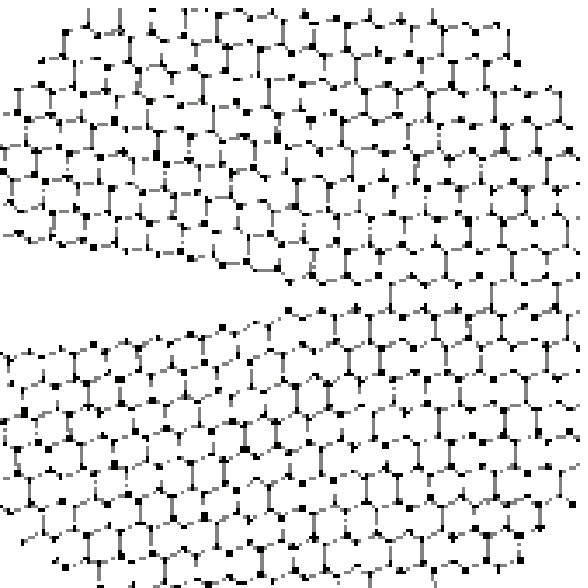, width = 1.4 in}}
\quad
\subfigure[$K=0.3068$]{\epsfig{file=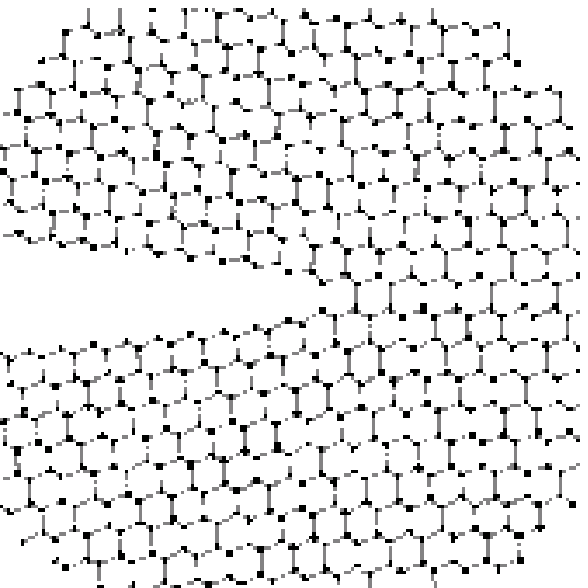, width = 1.4 in}}
\end{center}
\caption{\label{vizSW0} mSW-crack.}
\end{figure}

\subsection{\label{geometrySection}Geometry}

We simulated a cylindrical piece of silicon with a notch, making a `PacMan'
shape as in Fig.~\ref{notchschematic}(b), consisting of an inner \textit{core}
 region and an outer \textit{boundary region}. By focusing on just the
 initiation of fracture we avoid the need for large systems since we are not
 interested in the path the crack takes after the fracture (if we were, we 
would have a problem when the crack reached the edge of the core region and hit
 the boundary which is only a few lattice spacings away). We consider three
 notch geometries, which we call the $70^\circ$ (actually $70.5288^\circ$), 
$90^\circ$ and $125^\circ$ (actually $125.264^\circ$) geometries respectively,
 referring to the notch
 opening angles. The $70^\circ$ sample has \{111\} surfaces on the notch flanks 
and the plane of the sample is a \{110\} surface. The $90^\circ$ sample has 
\{110\} surfaces on the flanks and the plane is a \{100\} surface (in this case
 the crystal axes coincide with the coordinate axes). The $125^\circ$ sample has
a \{111\} on the bottom flank and a \{100\} surface on the top flank, while the
 plane of the sample is a \{110\} plane. In addition, we studied the zero 
degree notch geometry, corresponding to a standard crack. The crack plane is
 a \{111\} surface and is the $xz$ plane in the simulation, and the direction of
growth is the $\langle 211 \rangle$ direction, which is the $x$ direction in 
the simulation. The radius of the inner, core region in almost all the cases 
presented is 5 lattice spacings or about 27 \AA. The exceptions 
were the crack geometry for the EDIP potential (core radius was 7.5 lattice 
spacings---the larger size makes the ductile behavior of the potential more
 obvious) and the $90^\circ$ geometry with the MEAM potential (core radius was 
four lattice spacings because this potential is computationally more demanding).
 The coordinate system in each case is oriented so that the plane of
 the sample is the $xy$ plane and the notch is bisected by the $xz$ plane.

\begin{figure}
\begin{center}
\subfigure[$K=0.3$]{\epsfig{file=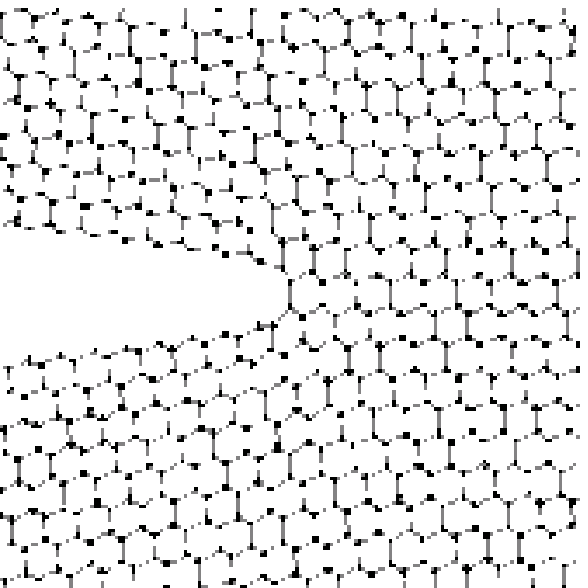,width=1.5 in}}
\hspace{0.1 in}
\subfigure[$K=0.4$]{\epsfig{file=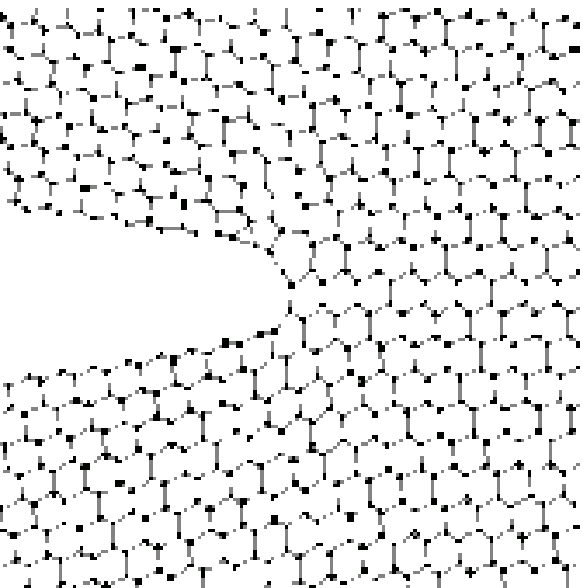,width=1.5 in}}

\vspace{0.2 in}

\subfigure[$K=0.56$]{\epsfig{file=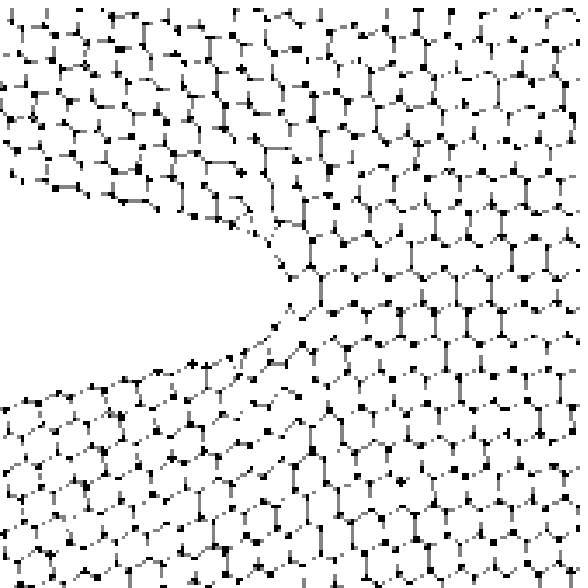,width=1.5 in}}
\hspace{0.1 in}
\subfigure[$K=0.6$]{\epsfig{file=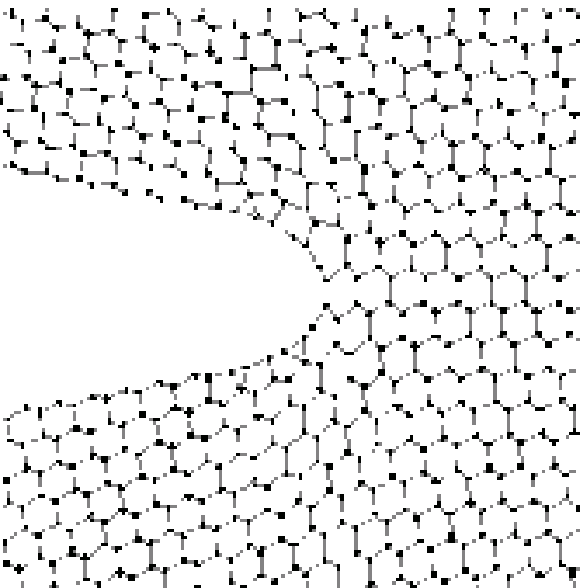,width=1.5 in}}

\vspace{0.2 in}

\subfigure[$K=0.66$]{\epsfig{file=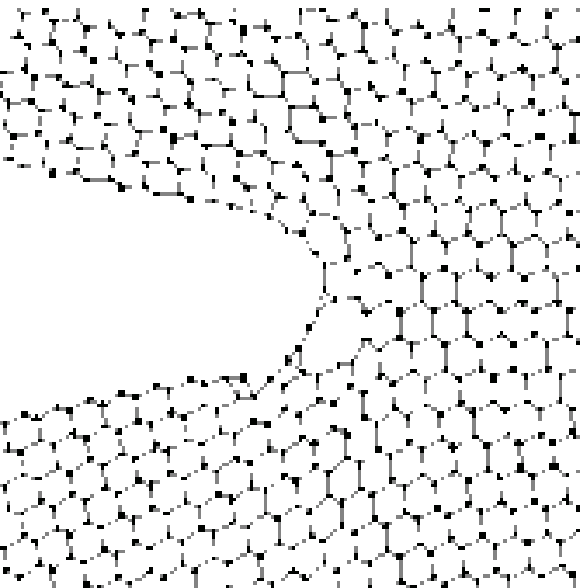,width=1.5 in}}
\hspace{0.1 in}
\subfigure[$K=0.76$]{\epsfig{file=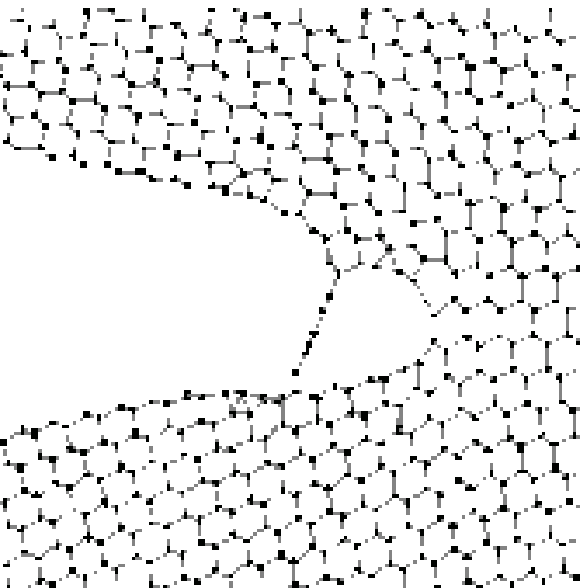,width=1.5 in}}
\caption{\label{vizEDIP0} EDIP-crack.}
\end{center}
\end{figure}

\subsection{\label{potentialsSection}Potentials }

We have used three different silicon potentials. The first is a modified 
form of the Stillinger-Weber \cite{Stillinger/Weber:1985} potential (mSW), in 
which the coefficient of the three body term has been multiplied by a factor of
 2. This has been noted by Hauch et al.\cite{Hauch/others:1999} to make the SW
 potential brittle; they were
 unable to obtain brittle fracture with the unmodified SW potential. However
it  worsens the likeness to real silicon in other respects such
 as melting point and elastic constants.\cite{Holland/Marder:1998, 
Holland/Marder:1999,Hauch/others:1999} The second potential is a more recent
 silicon potential known as ``environment-dependent interatomic potential'' 
(EDIP),\cite{Bazant/others:1997,Justo/others:1998} which is similar in form to
 SW but has an environmental dependence that makes it a many-body potential. 
Bernstein and coworkers\cite{Bernstein/others:1999,
Abraham/others:2000,Bernstein/Hess:2001} have used EDIP to simulate fracture
in silicon. They reported a fracture toughness about a
 factor of four too large when compared with experiment, and that fracture 
proceeds in a very ductile manner, accompanied by significant plastic 
deformation and disorder. On the other hand, using tight-binding molecular dynamics
 near the crack tip they successfully simulated brittle fracture in silicon. In 
view of the failure of many empirical potentials to simulate brittle fracture,
 P\'{e}rez and Gumbsch\cite{Perez/Gumbsch:2000}
 used density functional theory to simulate the fracture process, measuring 
lattice trapping barriers for different directions of crack growth on different
fracture planes. A reason for the failure of empirical potentials that has
been proposed in Ref.~\onlinecite{Bernstein/Hess:2001} is that their short-ranged 
nature necessarily requires large stresses to separate bonds. This however is
not the case in our third potential, which is the modified embedded atom method
(MEAM) of Baskes.\cite{Baskes:1992} This is a many-body potential similar to
the embedded atom method but with angular terms in the electron density; it has
 been fit to many elements including metals and semi-conductors. A significant 
feature of this potential is its use of ``three-body screening'' in addition 
to the usual pair cut-off distance. This means that atoms in the bulk see only their 
nearest neighbors, while surface atoms, on the other hand, can see any atoms 
above the surface (for example on the other side
of a crack) within the pair cut-off distance. The pair cut-off has been 
set to 6 \AA\ to allow the crack surfaces to see each 
other\cite{BaskesPrivate:2002} even after they have separated. The MEAM potential
has been used successfully to simulate dynamic fracture in 
silicon,\cite{BaskesPrivate:2002} and we have found it to be the most reliable
potential in our studies of notch fracture. In table~\ref{surfaceenergytable} we
list the low-index (relaxed, unreconstructed) surface energies for the three
 potentials.

\begin{figure}
\begin{center}
\subfigure[$K=0.183$]{\epsfig{file=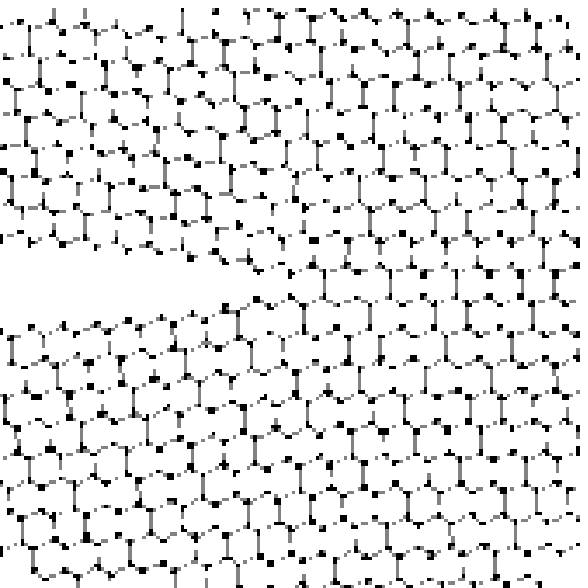,width=1.5 in}}
\hspace{0.1 in}
\subfigure[$K=0.1835$]{\epsfig{file=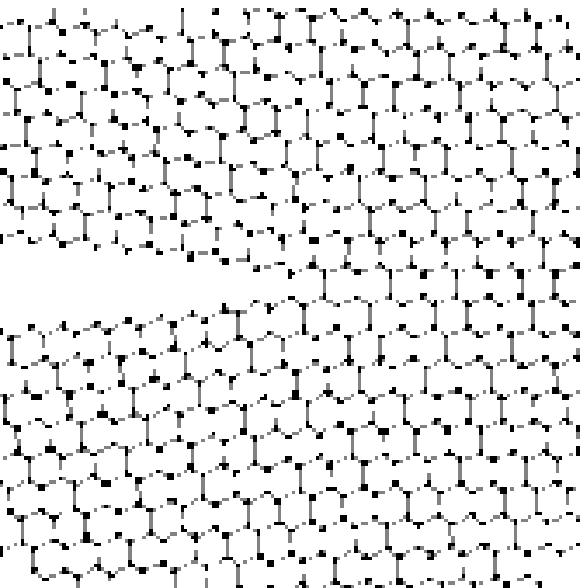,width=1.5 in}}
\caption{\label{vizMEAM0} MEAM-crack.}
\end{center}
\end{figure}

\subsection{\label{boundaryConditionsSection}Boundary Conditions}

The boundary conditions are as follows: in the $z$-direction (out of the page) 
there are periodic boundary conditions. The thickness of the sample in this
 direction is always one or two repeat distances of the lattice in that
 direction. For the $70^\circ$ and $125^\circ$ geometries the repeat distance
 is $\sqrt{2}a$ where $a$ is
 the cubic lattice constant; for the $90^\circ$ geometry it is $2a$. In the 
plane, the boundary conditions are that an layer of atoms
on the outside of the system has the positions given by the analytic formula
(\ref{analyticdisplacements}) for displacements from anisotropic linear
elasticity, with a specified stress intensity factor $K$. The thickness of the
layer is twice the cutoff distance of the potential, in order that the core
atoms feel properly surrounded by material.\footnote{It has to be twice because
of the three- and many-body terms in the potential.} We interpret the 
displacement formulas in terms of Eulerian coordinates, using an iterative
 procedure to compute the current positions. The numbers of core atoms were
890, 894, 1260, and 892, for the $0^\circ$ (crack), $70^\circ$, $90^\circ$ and
 $125^\circ$  systems respectively (except for the EDIP/crack case where the
core radius was
 7.5 lattice constants; there the number of core atoms was 2002). The number 
of boundary atoms depends on the
 potential (through the cutoff distance); it is typically about 500 atoms. For
the most part no special consideration was given to the lattice origin, which 
meant that by default it coincided with the notch tip.\footnote{When 
applying singular elastic deformations, a check ensures that an atom sitting at
 the location of the singularity is simply not displaced.} In a few cases it 
was necessary to shift the position of the origin in order to make sure that 
the notch flanks were made cleanly, in particular so that the \{111\} flanks
in the $70^\circ$ and $125^\circ$ geometries were complete close
packed \{111\} surfaces, rather than having dangling atoms.

\begin{figure}
\begin{center}
\subfigure[$K=0.3828$]{\epsfig{file=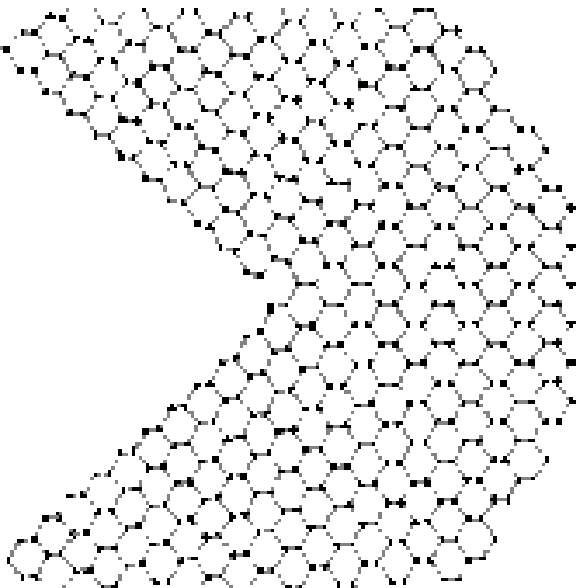,width=1.5 in}}
\hspace{0.1 in}
\subfigure[$K=0.3832$]{\epsfig{file=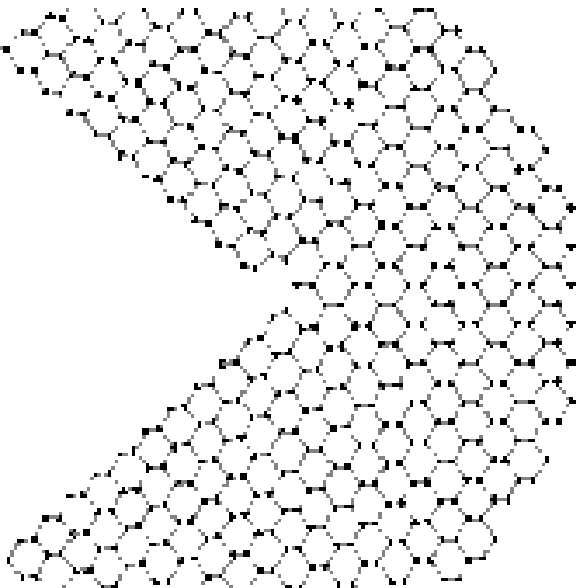,width=1.5 in}}

\subfigure[$K=0.4306$]{\epsfig{file=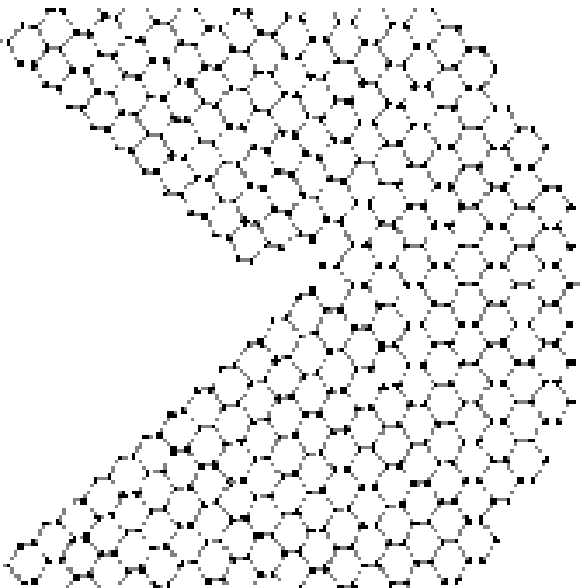,width=1.5 in}}

\caption{\label{vizSW70} mSW-$70^\circ$.}
\end{center}
\end{figure}

\subsection{\label{criticalStressIntensities}Critical stress intensities}

The simulation consists of alternating the following two
steps: (1) We increment the value of $K$ by a small amount, changing the
positions of the boundary atoms accordingly. (2) We
relax the interior atoms as follows. First we run about 50 steps of Langevin
molecular dynamics with a temperature of 500--600 K; the purpose of this is to 
break any symmetry (the $70^\circ$ and $90^\circ$ samples are symmetric about the
 $xz$ plane). It is still a zero temperature simulation; these
finite temperature steps are simply a way to introduce some noise. 
Next we run 500 time steps of the dynamical minimization technique known as
``MDmin'' (a Verlet time step is carried out, but after each velocity update,
 atoms whose velocities have negative dot-products with their forces have 
their velocities set to zero). Finally
500 time steps of conjugate gradients minimization are carried out. We
observe that the combination of both types of minimization is more effective
(converges to a zero force state more quickly) than either alone. The 
procedure generally results in the atoms having forces of around
 $10^{-5} eV/$\AA. 

The initial value for $K$ could be zero; however it turns out to be possible to
start from a fairly large value of $K$ by applying the analytic displacements
 to the whole system at first. When the critical $K$ value, $K_c$, is not yet known
the increment size is chosen reasonably large to quickly find the 
$K_c$. When this has been found, the simulation is restarted from a value below 
the critical value with smaller increments and a more accurate value for $K_c$ found.
 The increment is a measure of the uncertainty in $K_c$.

\section{Results}

\begin{figure}
\begin{center}
\subfigure[$K=0.2455$]{\epsfig{file=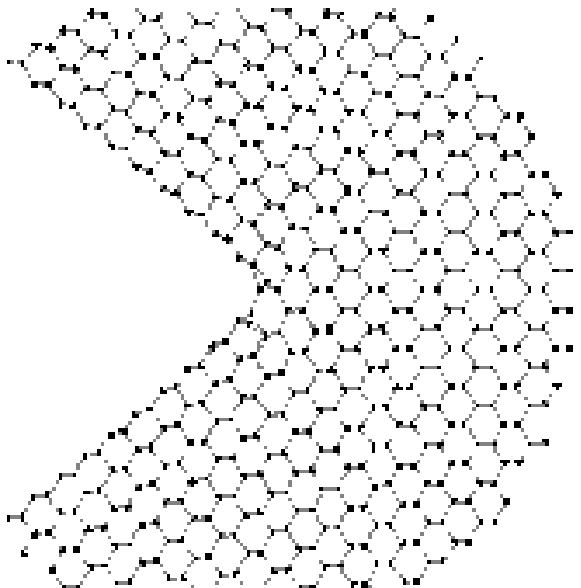,width=1.5 in}}
\hspace{0.1 in}
\subfigure[$K=0.2465$]{\epsfig{file=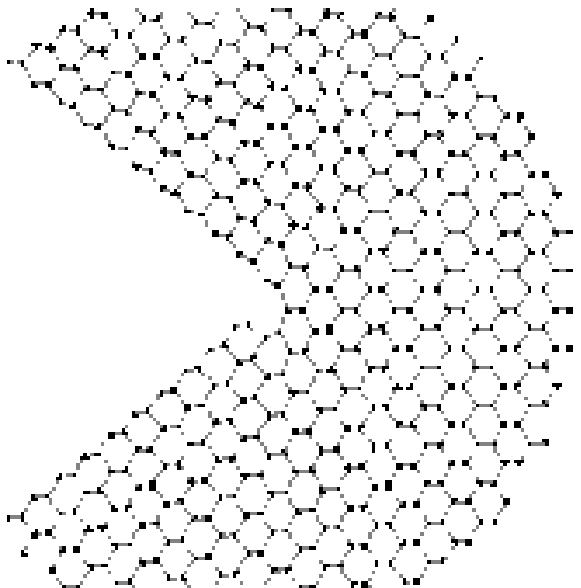,width=1.5 in}}
\caption{\label{vizEDIP70} EDIP-$70^\circ$.}
\end{center}
\end{figure}

\subsection{Observed fracture behavior}

\begin{figure}
\begin{center}
\subfigure[$K=0.266$]{\epsfig{file=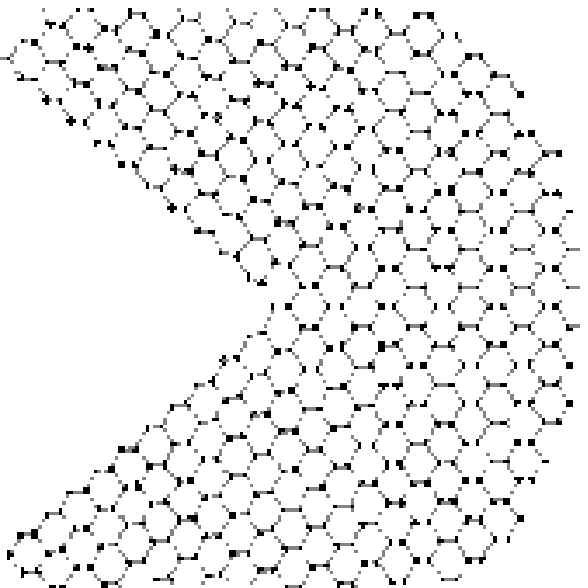,width=1.5 in}}
\hspace{0.1 in}
\subfigure[$K=0.268$]{\epsfig{file=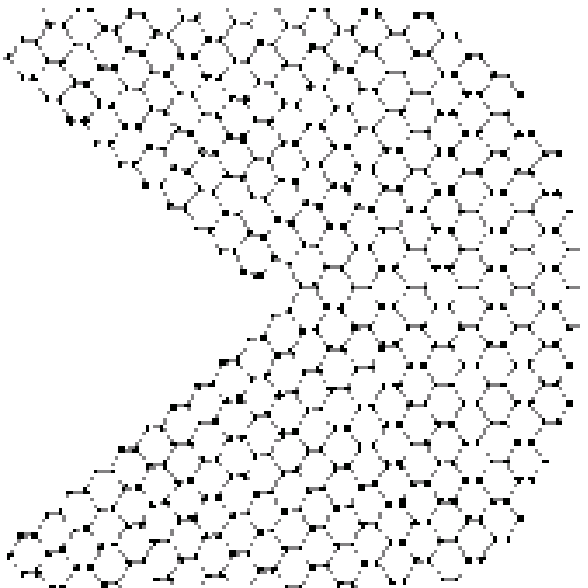,width=1.5 in}}

\caption{\label{vizMEAM70} MEAM-$70^\circ$.}
\end{center}
\end{figure}

\begin{figure}
\begin{center}
\subfigure[$K=0.3860$]{\epsfig{file=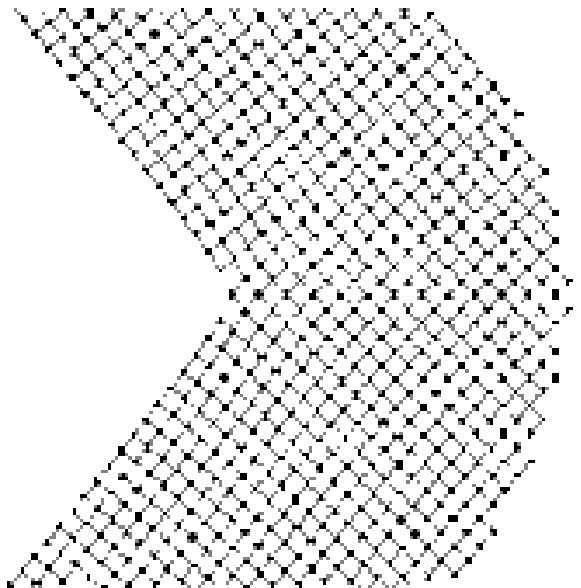,width=1.5 in}}
\hspace{0.1 in}
\subfigure[$K=0.3863$]{\epsfig{file=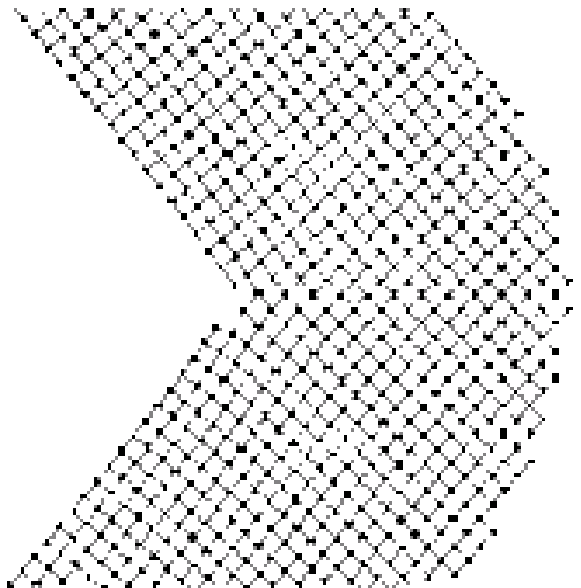,width=1.5 in}}

\subfigure[$K=0.3871$]{\epsfig{file=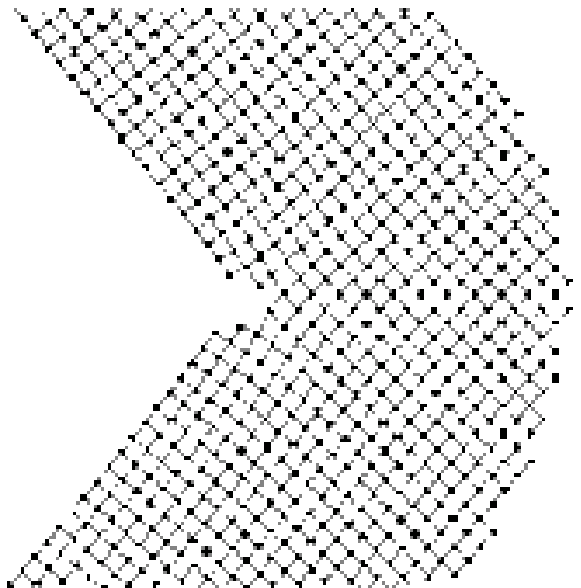,width=1.5 in}}

\caption{\label{vizSW90} mSW-$90^\circ$.}
\end{center}
\end{figure}

\begin{figure}
\begin{center}
\subfigure[$K=0.451$]{\epsfig{file=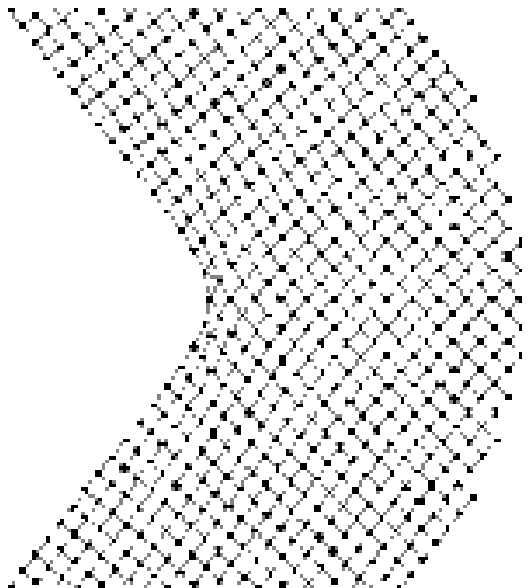,width=1.5 in}}
\hspace{0.1 in}
\subfigure[$K=0.491$]{\epsfig{file=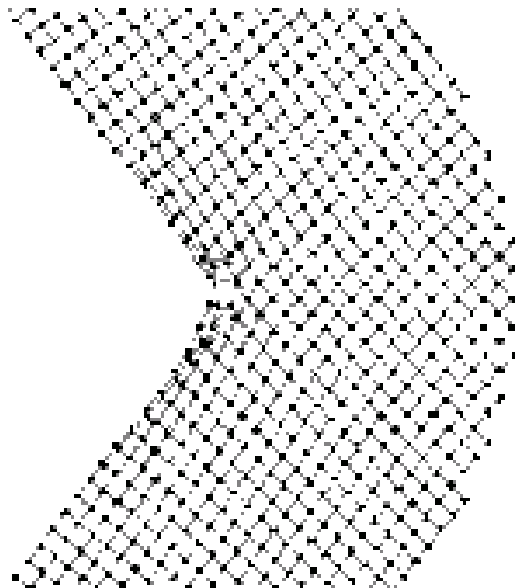,width=1.5 in}}

\vspace{0.2 in}

\subfigure[$K=0.511$]{\epsfig{file=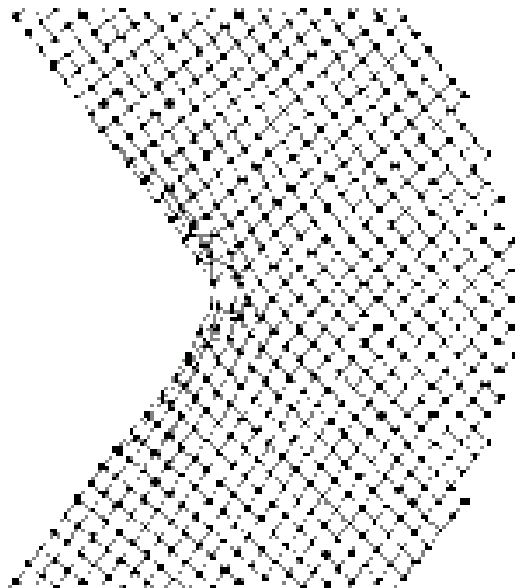,width=1.5 in}}
\hspace{0.1 in}
\subfigure[$K=0.521$]{\epsfig{file=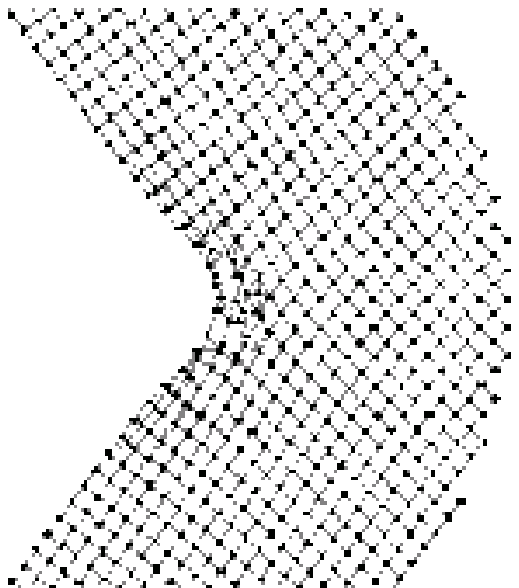,width=1.5 in}}

\vspace{0.2 in}

\subfigure[$K=0.531$]{\epsfig{file=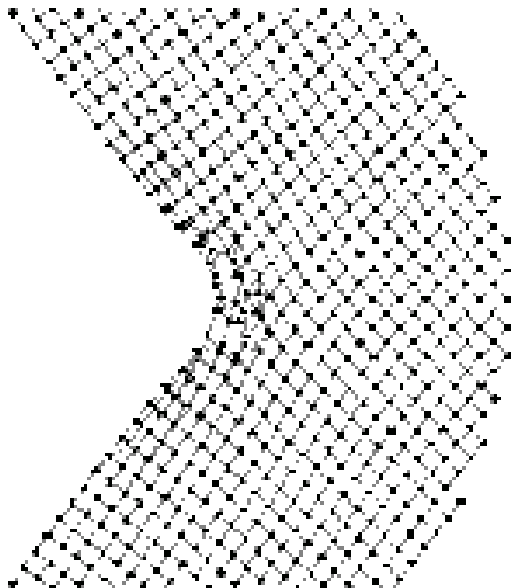,width=1.5 in}}
\hspace{0.1 in}
\subfigure[$K=0.6$]{\epsfig{file=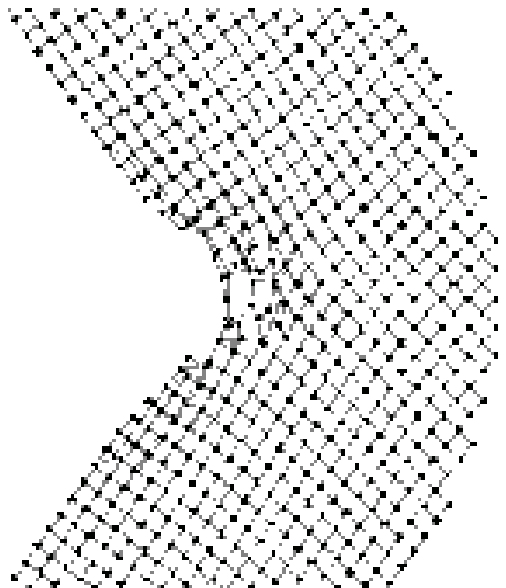,width=1.5 in}}

\caption{\label{vizEDIP90} EDIP-$90^\circ$.}
\end{center}
\end{figure}

\begin{figure}
\begin{center}
\subfigure[$K=0.293$]{\epsfig{file=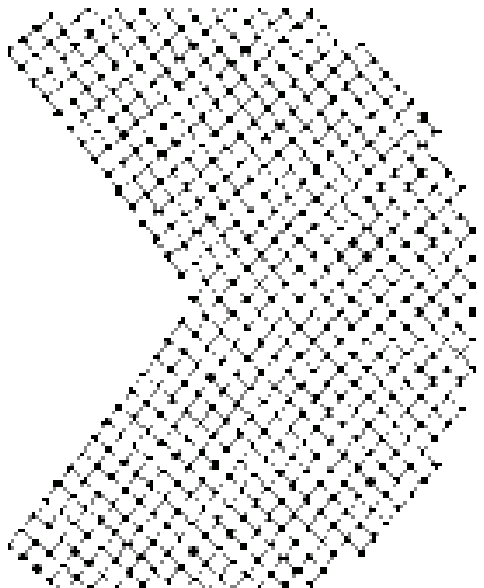,width=1.5 in}}
\hspace{0.1 in}
\subfigure[$K=0.2935$]{\epsfig{file=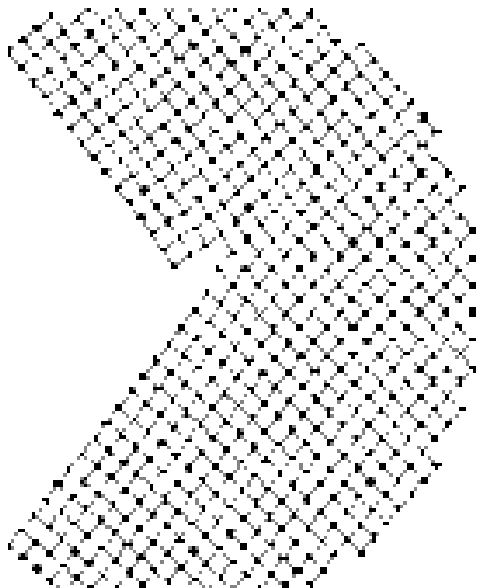,width=1.5 in}}

\subfigure[$K=0.34$]{\epsfig{file=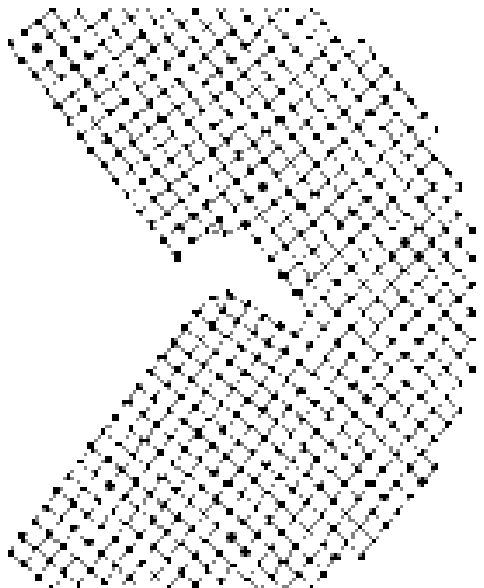,width=1.5 in}}

\caption{\label{vizMEAM90} MEAM-$90^\circ$.}
\end{center}
\end{figure}

\begin{figure}
\begin{center}
\subfigure[$K=0.3302$]{\epsfig{file=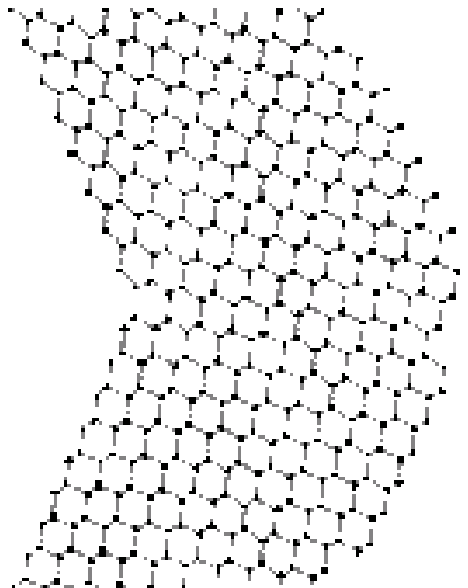,width=1.5 in}}
\hspace{0.1 in}
\subfigure[$K=0.3304$]{\epsfig{file=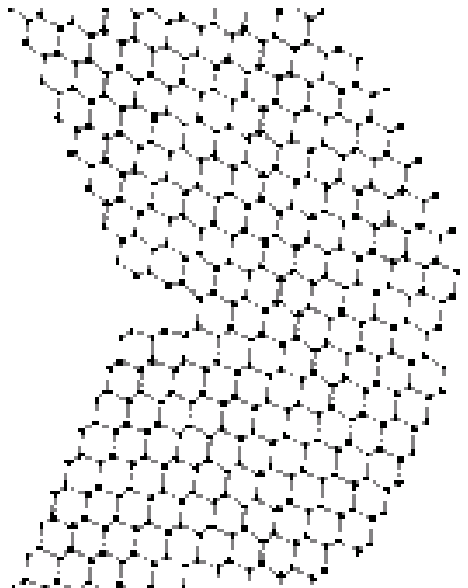,width=1.5 in}}

\subfigure[$K=0.3323$]{\epsfig{file=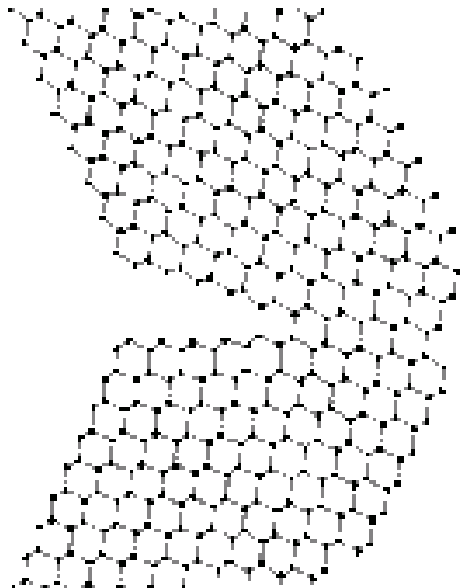,width=1.5 in}}

\caption{\label{vizSW125} mSW-$125^\circ$.}
\end{center}
\end{figure}

\begin{figure}
\begin{center}
\subfigure[$K=0.43$]{\epsfig{file=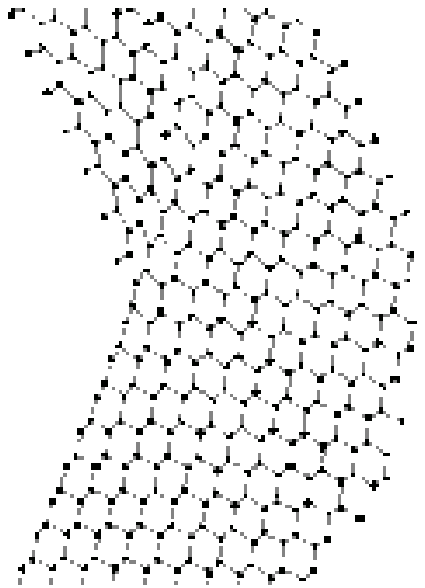,width=1.5 in}}
\hspace{0.1 in}
\subfigure[$K=0.44$]{\epsfig{file=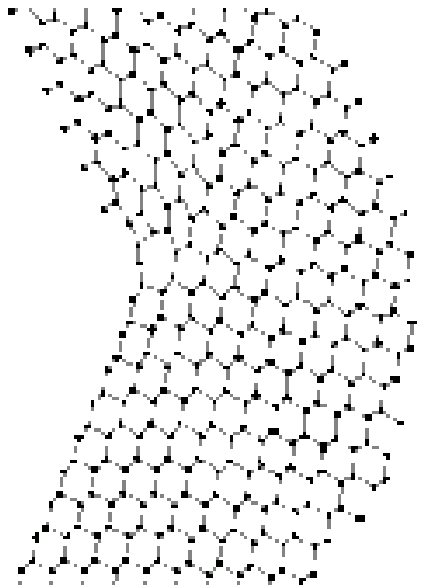,width=1.5 in}}

\vspace{0.2 in}

\subfigure[$K=0.52$]{\epsfig{file=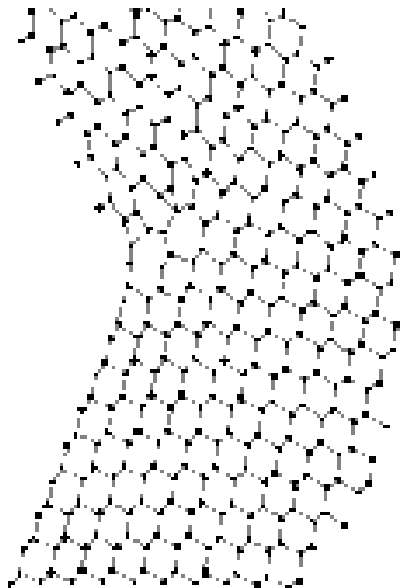,width=1.5 in}}
\hspace{0.1 in}
\subfigure[$K=0.57$]{\epsfig{file=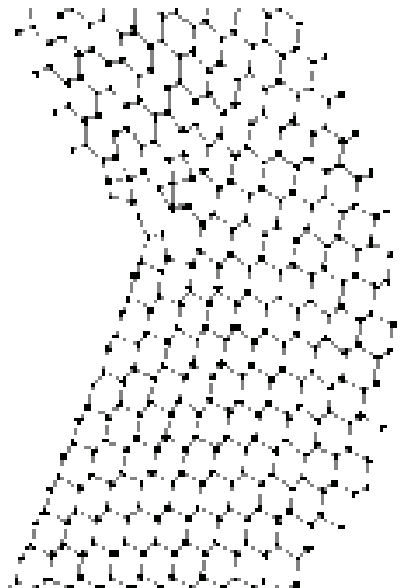,width=1.5 in}}

\vspace{0.2 in}

\subfigure[$K=0.58$]{\epsfig{file=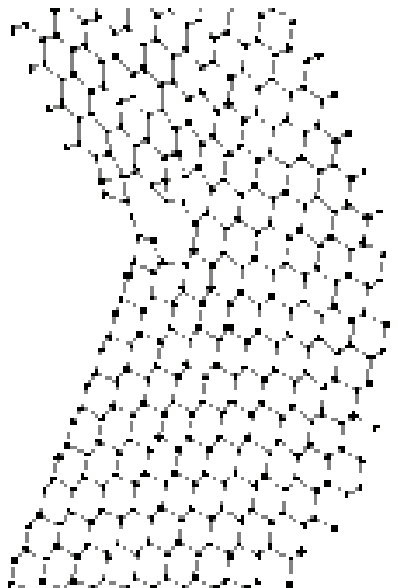,width=1.5 in}}
\hspace{0.1 in}
\subfigure[$K=0.59$]{\epsfig{file=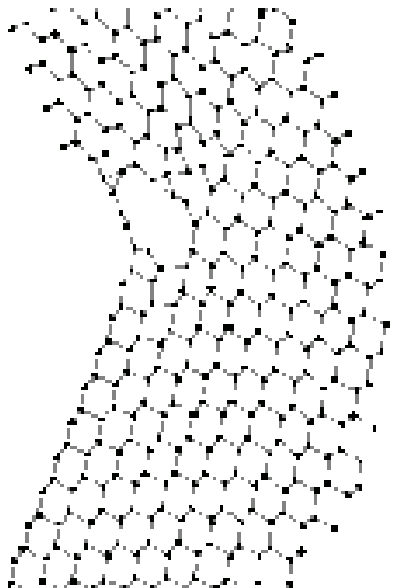,width=1.5 in}}

\caption{\label{vizEDIP125} EDIP-$125^\circ$.}
\end{center}
\end{figure}

\begin{figure}
\begin{center}
\subfigure[$K=0.2195$]{\epsfig{file=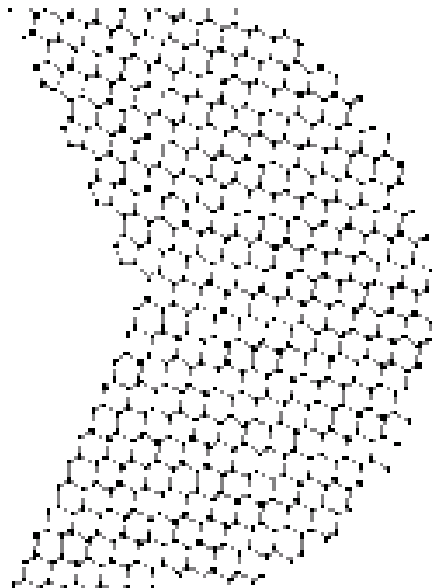,width=1.5 in}}
\hspace{0.1 in}
\subfigure[$K=0.22$]{\epsfig{file=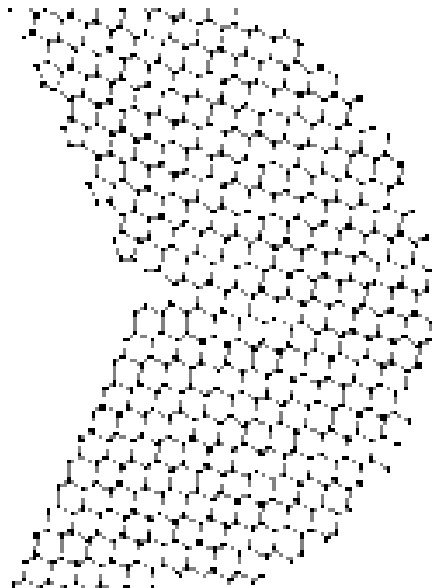,width=1.5 in}}

\caption{\label{vizMEAM125} MEAM-$125^\circ$.}
\end{center}
\end{figure}

We observe brittle cleavage of the simulated crystals at definite values of $K$
for all geometries using the mSW and MEAM potentials, but only for the $70^\circ$
geometry when using the EDIP potential.
Figures \ref{vizSW0}--\ref{vizMEAM125} show snapshots of the 
simulation process for the different geometries and potentials. In most cases
two or three snapshots are shown: one of the configuration immediately before
crack initiation, one of the configuration immediately after initiation, and 
possibly one
of a ``late-stage'' configuration, to illustrate the fracture plane more
vividly; generally this was chosen to be the configuration corresponding to the
highest applied load, which depended on how long the 
simulation was run past the initiation point. For the EDIP potential, which
 gives unphysical ductile behavior 
(except in one case, the $70^\circ$ geometry), more snapshots are shown, in 
order to illustrate the plastic behavior more completely, since a variety of
stages is involved.

The behavior in crack geometries is shown in 
Figs.~\ref{vizSW0}--\ref{vizMEAM0}. The initial applied load must be such that 
no crack healing takes place upon relaxation, so that the location of the crack
corresponds to the center of the system (in reference to which 
the boundary displacements are calculated). In this case we are not 
investigating crack initiation (since the notch is already a crack)but crack 
growth; the critical $K_c$ is
defined as that at which the crack advances, or when the next bond across the
crack plane breaks. This is somewhat hard to see in the figures; one must
count atoms along the crack surface and compare from one figure to another to
see that growth has occurred.

The mSW and MEAM potentials produce similar, brittle, fracture behavior. 
The EDIP potential produces quite different behavior; the crack propagates in
 a ductile manner. Frame~(a) shows the configuration
before any plastic deformation has taken place. Frame~(b) shows what appears to
be the nucleation of a dislocation onto the \{110\} slip plane which is at an
 angle of $54.6^\circ$ to the positive $x$-axis. By frame (c) the crack tip
has blunted noticeably, and in frame~(d) a growth of the blunted crack by about
 a lattice constant has taken place---we take the stress intensity at this 
stage
to be the critical value. Frames~(e) and (f) show a void nucleating and growing
behind the crack tip, which would under further loading join with the
 crack---coalescence of voids the essence of ductile crack growth.

In the $70^\circ$ system (Figs.~\ref{vizSW70}--\ref{vizMEAM70}) fracture 
occurs along a \{111\} plane. There are
 two choices for this, symmetrically placed with respect to the $xz$ plane.
Here all three potentials produced brittle behavior; this was the only geometry
in which the EDIP potential did so. Possible reasons for
this exception are 
discussed in section \ref{discussion}. However, when the origin was not shifted as
 mentioned in section \ref{criticalStressIntensities}, so that the notch flanks
had dangling atoms, the EDIP-behavior was quite different: the notch 
blunted to a width of several atomic spacings.

The behavior for $90^\circ$ models is shown in 
Figs.~\ref{vizSW90}--\ref{vizMEAM90}. We observe three different behaviors for 
three different potentials---providing a cautionary demonstration of the 
limitations of empirical potentials. The easy cleavage planes available here 
are the \{110\}
planes which are extensions of the notch flanks. The mSW model starts to cleave
along the lower of these (the extension of the upper flank) but the crack advances
only one atomic before cleavage switches to an adjacent parallel plane. The
net result is a kind of ``unzipping'' along the hard \{100\} plane.  This is 
presumably because the peak in the normal stress across this plane, compared
to the normal stress at the $45^\circ$ angle, outweighs the increased cost of
cleavage (but note that the surface energy ratio $\gamma_{100}/ \gamma_{111}$ is in fact 
lower for MEAM, which cleaves on the \{110\} plane---see table 
\ref{surfaceenergytable} for the energies of the different
 surfaces according to the different potentials). The EDIP potential
deforms plastically in this case, as depicted in the six frames of 
Fig.~\ref{vizEDIP90}. It is harder to identify specific processes here than in
the $70^\circ$ case, including where crack growth starts, though it seems to 
have definitely started by the frame (c)($K_c = 0.6$). The MEAM potential
behaves in the manner most consistent with experiment, namely cleaving on
\{110\} planes, and switching from one to the other---this is illustrated 
dramatically in the third frame of Fig.~\ref{vizMEAM90}. Experimentally,
switching between planes, when it happens, occurs over longer length
 scales ($25 \mu m$ for the $70^\circ$ case\cite{Suwito/Dunn/Cunningham:1998}),
 although
the behavior at atomic length scales has not been examined. Too much should
not be read into the switching we observe, because once cleavage has occurred
over such distances the proximity of the boundary probably has a large effect
on the effective driving force on the crack.

For the $125^\circ$ geometry (Figs.~\ref{vizSW125}--\ref{vizMEAM125}), there 
are again two \{111\} planes to choose from but they are not symmetrically 
placed. Fracture occurs for the mSW and MEAM potentials on the one closest to 
the $xz$ plane, i.e., closest to the plane of maximum normal stress, which is 
the $(11\bar 1)$ plane. The direction of growth is $[2 1 \bar 1]$,
 and growth proceeds much more readily than in the other notch geometries, 
presumably because it is almost along the maximum stress plane. In the EDIP
 system, 
plastic deformation is favored over cleavage. This appears to proceed as 
follows: First, slip occurs on the $(11\bar 1)$ plane in the $[2 1
 \bar 1]$ direction, as a single edge dislocation is nucleated
 (frame~(a)--frame~(b)). 
Next, slip occurs on the other \{111\} plane, the $(1\bar 1 1)$ plane, in the 
$[2 1 \bar 1]$ direction, with two dislocations being nucleated 
(frame~(b)--frame~(c)--frame~(d)), on adjacent $(1\bar 1 1)$ planes. In the 
last two frames a void appears and grows.

\subsection{Critical stress intensities}

\begin{figure}
\begin{center}
\subfigure[Units based on $eV$
and \AA\ ]{\epsfig{file=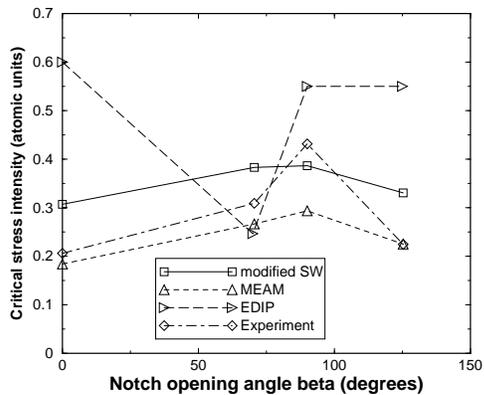,width = 2.5 in}}

\vspace{0.3 in}

\subfigure[SI units]{\epsfig{file=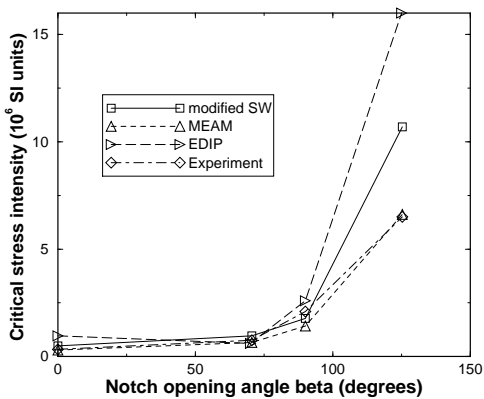,width = 2.5 in}}
\end{center}
\caption{\label{KvsAnglePlots}Computed critical stress intensities for the three potentials and experiment.}
\end{figure}

The values of $K_c$, for the different potentials as well as from experiment,
 are listed in table \ref{criticalKtable} and plotted
 in Fig.~\ref{KvsAnglePlots}. The increment size for $K$ is listed as an 
estimate of the error in $K_c$. The values for ductile
fracture from the EDIP potential are marked with an asterisk as a 
reminder that the definition of $K_c$ in these cases is problematic. The
 experimental value for the crack geometry is from 
Ref.~\onlinecite{Chen/Leipold:1980}.  Notice that the critical stress
 intensities
 for difference angles are almost the same in atomic
 units, and differ by more than a factor of ten in standard 
units\footnote{Note
 that the exact conversion factor depends on the eigenvalue $\lambda$ which 
depends on the potential (see appendix \ref{unitsConversions}), but for a 
given angle the dependence on potential is quite small.} To check for finite 
size effects, we repeated the measurement on the $70^\circ$
geometry, but with larger radius of 8 \AA, using the MEAM potential. In this
case $K_c$ was determined to be $0.262\pm 0.001$, or about $1.7$\% lower than
 the value from the smaller system. This indicates that finite size effects are
 small, but not negligible. To compensate for them without using larger systems
 a flexible boundary method could be used, involving higher order
``multipoles'' of the elastic field, appropriate for the notch (i.e., solutions
 with $\lambda < 0 $), which could be relaxed.

\begin{table}
\caption{ \label{criticalKtable}Critical stress intensity values for different
 geometries and potentials, including experimental data from
 Refs.~\onlinecite{Suwito/Dunn/Cunningham:1998, Suwito/others:1999}.}
\begin{center}
\begin{tabular}{|l|l|l|l|l|l|}
\hline
Potential & Geom. & $K_c$ & Error & Griffith & $K_c$ (SI) \\
\hline
mSW   & 0  & 0.3068 & 0.00036 & 0.19509 &  $4.9 \times 10^5$\\ 
mSW   & 70  & 0.3832 & 0.00036 & - & $9.6\times10^5$ \\ 
mSW   & 90  & 0.3863 & 0.00035 & - & $1.78\times10^6$ \\ 
mSW   & 125 & 0.3304 & 0.00016 & - & $1.07\times10^7$ \\ 
EDIP & 0  & 0.6* & 0.02 & 0.14634 & $9.6\times10^5$ \\  
EDIP & 70  & 0.2465 & 0.001 & - & $6.1 \times 10^5$\\ 
EDIP & 90  & 0.5--0.6* & 0.0005  & - & $2.4$--$2.8 \times 10^6$ \\ 
EDIP & 125 & 0.5--0.6* & 0.001 & - & $1.5$--$1.8 \times 10^7$ \\ 
MEAM & 0   & 0.184 & 0.0005 & 0.16406 & $3 \times 10^5$ \\  
MEAM & 70  & 0.2665 & 0.0005 & - &  $6.57 \times 10^5$ \\ 
MEAM & 90  & 0.2935 & 0.0005 & - & $1.42 \times 10^6$ \\ 
MEAM & 125 & 0.22  & 0.0005  & - &  $6.47 \times 10^6$ \\
Expt & 0 & 0.2060 & - & 0.1776 & $3.3 \times 10^5$ \\
Expt & 70  & 0.31 & 10\% & - & $7.6\times10^5$ \\
Expt & 90  & 0.43 & 10\% & - & $2.1\times10^6$ \\
Expt & 125 & 0.22 & 10\% & - & $6.5\times10^6$\\
\hline
\end{tabular}
\end{center}
\end{table}

\section{\label{discussion} Discussion}

\subsection{Critical stress intensities}

Comparisons are easier to make when looking at the data plotted using atomic
scale units. Then the data for the two brittle potentials is a gentle, almost 
horizontal, curve. The experimental data mostly lies between that for the
MEAM potential and that for the mSW potential, but significantly closer
to the former. The exception is the $90^\circ$ case where the experimental 
value jumps to higher than the mSW value. Since the curves from the two
 potentials are very similar in shape---the main difference seems to be an
 overall shift or factor---and the jump in the experimental value at $90^\circ$
is a departure from this shape, it would not be meaningful to assert that the
mSW potential does a better job in predicting $K_c$ in the $90^\circ$ case. 
For the other angles the MEAM values are more or less within experimental error
of experiment: the error (standard deviation across all the tested samples) is 
close to 10\% in all cases (the error is not available for the crack case), and
 the
percentage differences of the MEAM values with respect to the experimental
 values are $-10$\%, $-14$\%, $-32$\% and $-0.5$\% for the $0^\circ$, 
$70^\circ$, 
$90^\circ$ and $125^\circ$ geometries respectively. The 0.5\% is clearly
 fortuitous. Note that the experimental error bar is not enough to account for 
the anomalously high value for the $90^\circ$ case; there must be some feature
 of the physics or energetics of fracture initiation in this geometry that is
missing from the others, and missing from the simulation.

An interesting question is why the EDIP potential behaves unlike the other 
potentials and experiment except at one particular geometry, the 
$70^\circ$ one. Possibly there is some feature of this geometry that suppresses
the nucleation of dislocations. Dislocation Burgers vectors in silicon are
 always in  $\half\langle 110 \rangle$ directions, since these are the shortest
 perfect lattice vectors in the diamond lattice.\cite{Hirth/Lothe:1982} The 
periodic boundary conditions constrain possible dislocation lines to be out of
 the plane. Moreover, since we are considering only mode I and II loading, we
expect slip to be within the plane, so we consider only edge dislocations. In
the $70^\circ$ geometry the $\half\langle 110 \rangle$ direction that is
 available within the plane is at an angle of $90^\circ$ to the $x$-axis, while the
cleavage plane is at an angle of $35.26^\circ$. Looking at 
Fig.~\ref{normalShearAngular}, we can see that the shear stress and normal 
stresses on these planes respectively are both near their maximum values, 
although the ratio of shear stress to normal stress is 0.43 (both plots are
 normalized with respect to $K$ and $r$.). Without knowing the 
values of stress needed to initiate slip or cleavage on these respective
 planes, this ratio is not enough to explain anything. What we can do is 
compare the same ratio for the other geometries and check if it is higher in
 other cases, thereby tending to make slip more favorable than cleavage, for
a given potential (namely, EDIP). The numbers---angles for slip and cleavage
 and the appropriate stress ratio---are shown in 
table~\ref{slipCrackRatios}. Unfortunately, there is no conclusive trend. The
ratio for $90^\circ$ is indeed higher than that for $70^\circ$ but the others
are lower, and EDIP notches suffer plastic deformation in all of the other
 cases.

\begin{table}
\caption{\label{slipCrackRatios}Angles of slip planes and crack planes and 
ratio of shear to normal stress for different geometries.}
\begin{center}
\begin{tabular}{|l|l|l|l|}
\hline
geometry & slip plane  & crack plane & shear/normal \\
\hline
70   & $90^\circ$ & $35.26^\circ$ & 0.43 \\
90   & $45^\circ$ & $45^\circ$ & 0.52 \\
125  & $27.34^\circ$ & $-7.90^\circ$ &  0.34\\
0    & $54.6 ^\circ$ & $0^\circ$ & 0.37 \\
\hline
\end{tabular}
\end{center}
\end{table}

For the crack cases we can make a comparison of our results with the so-called 
\textit{Griffith criterion} for crack propagation. This comes from setting the
 energy release rate equal to twice the surface energy. An expression for the 
mode I energy release rate in terms of the stress intensity factor is given in 
Ref.~\onlinecite{Sih/Paris/Irwin:1965}; setting it equal to twice the surface 
energy leads to the following expression for the critical stress intensity 
factor

\begin{equation}
\bfa{K}_{\text{Griffith}} = \left( \frac{2 \gamma}{\pi b_{22}\text{Im}((\mu_1+\mu_2)
/(\mu_1 \mu_2))}  \right)^{\half}
\end{equation}

\nod where $\mu_1$ and $\mu_2$ are the roots of a characteristic polynomial
which depends on the elastic constants  and $b_{22}$ is an element of the 
compliance tensor
 for plane strain. The ratio $K_c/K_{\text{Griffith}}$ is associated with 
lattice trapping, when fracture is brittle. This ratio is 1.57 for the mSW 
potential and 1.12 for MEAM. These values are respectively somewhat larger and
 somewhat smaller than the ratio 1.25 determined by  P\'{e}rez and Gumbsch 
using total energy pseudopotential calculations\cite{Perez/Gumbsch:2000} (our
 $K_c$ corresponds to their $K_+$). In 
the EDIP case, where fracture proceeds only accompanied by significant plastic
 deformation, $K_c$ is four times the Griffith value.

In our simulations, for a given potential, only one fracture behavior is
 observed, in contrast to what was observed in the experiments of Suwito et
 al.\cite{Suwito/Dunn/Cunningham:1998} Specifically, in the case of the 
$70^\circ$ geometry, they observed three different ``modes'' (not to be
 confused with loading modes), including propagation on the (110) plane,
 yet we have observed  cleavage only on \{111\} planes in this geometry.
 It is possible that finite temperature, and the relative heights of different
 lattice trapping barriers, play an important role here. More likely it is 
related to experimental microcracks or defects near the crack tip. In any case,
it would be of great benefit to systematically calculate the barriers for 
different processes that can occur at a notch (or crack) tip, as a function
 of applied load.

A further point to note, and a warning, is this: In comparing simulations 
involving such very small length scales (27 \AA)
 to experiment it is appropriate to consider the question of whether the 
experimental notches are indeed as sharp as we have made our simulated notches.
Suwito et al.\cite{Suwito/Dunn/Cunningham:1998} could only put an upper limit
of $0.8 \mu m$ on the radius of curvature of their notches, although notch 
radii of the order of $10nm$ have been reported in etched 
silicon.\footnote{This fact is mentioned without any citation in 
Ref.~\onlinecite{Suwito/others:1999}.} The addition of just a few atoms right at the
notch tip would presumably have a significant effect on the energetics of 
cleavage initiation. We have not made any investigation of this, and this 
question should be borne in mind given the absence of experimental data 
characterizing the notch tip at the atomic scale. Nevertheless, the 
success of our simulations provides an important indication that these notches
 are indeed atomistically sharp.

\section{Summary}\label{summarySilicon}

We have determined by atomistic simulation the critical stress intensities to
initiate fracture in notched single crystal silicon samples. The samples had
angles of $0^\circ$ (a crack), $70.5233^\circ$, $90^\circ$ and 
$125.264^\circ$---chosen so that the flanks of the notches were low index 
crystal
 planes. These geometries correspond to those studied experimentally in 
measurements of critical stress intensities for fracture initiation. Of the
three potentials used, modified Stillinger-Weber (mSW), environment-dependent 
interatomic potential (EDIP) and modified embedded atom method (MEAM), MEAM
produced the most realistic behavior. The mSW potential produced brittle fracture,
 but its
resemblance to silicon in other respects is quite weak. Except in the case
of the $70^\circ$ notch, the EDIP potential gives ductile fracture
with a critical stress intensity factor, which
is much higher than that determined using the other potentials, or by 
experiment.

\section{Acknowledgments}

We thank Zhiliang Zhang for inspiration and helpful discussions, and Noam
Bernstein for helpful discussions. We also thank Mike Baskes for help in coding
the MEAM potential. This work was
financed by NSF-KDI grant No. 9873214 and NSF-ITR grant No. ACI-0085969. Atomic
 position visualization figures were produced using the \textsc{dan} program,
 developed by N. Bernstein at Harvard University and the Naval Research 
Laboratory.

\appendix

\section{\label{unitsConversions}Units and Conversions}

Three different sets of units are used in this paper. To each atomic potential
(Stillinger-Weber, EDIP, MEAM) is associated a set of atomic units (EDIP and 
MEAM use the same units); also we often wish to use SI units to
compare to experiment. In the context of this paper there is the further
subtlety that the units of the chief quantity under consideration, namely the
stress intensity factor $K$, are not simple powers of base units but involve a
non-trivial exponent $\lambda$ which is a function of geometry and potential.
 In fact the SI units units for $K$ are $Pa\, m^{1-\lambda}$ which for brevity
 we simply refer to as `standard units' in the paper.

The units for an atomic potential are determined by specifying the unit of
energy and that of length (for dynamics the unit of time is determined from
these and the particle mass). The SW potential as originally written down did
 not have units built into it. By taking the energy unit to be $\epsilon = 
2.1672 eV=3.4723\times 10^{-19}J$  and the length unit to be $\sigma = 2.0951$
\AA, the authors modeled molten silicon.\cite{Stillinger/Weber:1985} However
 other authors\cite{Balamane/others:1992} have taken the
 energy unit to  be $\epsilon = 2.315 eV$. The difference is not really 
important since we have modified the potential itself to make it more brittle
 so the resemblance to real silicon is reduced noticeably. When expressing
 quantities in terms of $eV-$\AA units we use the second
 scaling which is more common. The EDIP and MEAM
 potentials have $\epsilon=1eV$ and $\sigma=1$\AA built in as their
 units. Since $\sigma \sim K r^{\lambda - 1}$, the units of $K$ are
$[stress]/[length]^{\lambda-1} = [energy]/[length]^{2+\lambda}$, so to convert 
a value for $K$ in atomic units to SI units, one uses the conversion factor
$\epsilon/\sigma^{2+\lambda}$. Table \ref{conversionfactortable} gives the
factors for the three potentials and the geometries studied in this
paper.

\begin{table}[h]
\caption{\label{conversionfactortable}Unit conversion factors for $K$.}
\begin{center}
\begin{tabular}{|c|c|l|r|}
\hline
Potential & geometry & $\lambda$ & factor \\
\hline
mSW   & 0   & 0.5 & 1602000 \\ 
mSW   & 70  & 0.51954 & 2510000 \\ 
mSW   & 90  & 0.54597 & 4620000 \\ 
mSW   & 125 & 0.63047 & 32320000 \\ 
EDIP & 0  & 0.5 & 1602000  \\
EDIP & 70  & 0.51922 & 2490000 \\
EDIP & 90  & 0.54708 & 4730000   \\
EDIP & 125 & 0.62844 & 30840000  \\
MEAM & 0  & 0.5 & 1602000  \\
MEAM & 70  & 0.51875 & 2467000 \\
MEAM & 90  & 0.54794 & 4832000   \\
MEAM & 125 & 0.62639 & 29420000  \\
\hline
\end{tabular}
\end{center}
\end{table}


\section{Stroh formalism for notches}\label{strohappendix}

Here we summarize the application of the Stroh formalism to the notch
problem. More details are available in 
Refs.~\onlinecite{Suwito/Dunn/Cunningham:1998, Suwito/others:1999, 
Labossiere/Dunn:1998}. We can write the solution for the 
displacement field $\bfa{u}$ and the stress function $\phi$ as 

\begin{equation}
\bfa{u} = \sum_{\alpha=1}^6 \bfa{a}_\alpha f_\alpha(z_\alpha)
\end{equation}

\begin{equation}
\phi = \sum_{\alpha=1}^6 \bfa{b}_\alpha f_\alpha(z_\alpha)
\end{equation}

\nod The independent variable here is the complex variable
 $z_\alpha = x_1 + p_\alpha
x_2$. The stress function $\phi$ determines the stresses through $\sigma_{i1} =
-\phi_{i,2}$ and $\sigma_{i2} = \phi_{i,1}$. The $p_\alpha, \bfa{a}_\alpha$ and
$\bfa{b}_\alpha$ come from solving the following eigenvalue problem:

\begin{equation}
(\bfa{Q} + p(\bfa{R+R^T}) + p^2 \bfa{T}) \bfa{a} =  0
\end{equation}

\nod where

\begin{equation}
\begin{split}
\bfa{Q} &= 
\left[
\begin{array}{ccc}
C_{11} & C_{16} & C_{15} \\
C_{16}   & C_{66} & C_{56} \\
C_{15}   & C_{56} & C_{55}
\end{array}
\right]
\bfa{R} =
\left[
\begin{array}{ccc}
C_{16} & C_{12} & C_{14} \\
C_{66} & C_{26} & C_{46} \\
C_{56} & C_{25} & C_{45}
\end{array}  
\right]
 \\
\bfa{T} &= 
\left[
\begin{array}{ccc}
C_{66} & C_{26} & C_{46} \\
C_{26} & C_{22} & C_{24} \\
C_{46} & C_{24} & C_{44}
\end{array}
\right]
\end{split}
\end{equation}

\nod The above is general within the context of two-dimensional anisotropic
elasticity. To specify the notch problem we choose a form of the arbitrary
function $f$ to which we can apply the boundary conditions of the 
problem---that notch flanks are traction-free. The following choice does the
 trick:

\begin{equation}
f_\alpha(z_\alpha) = \frac{1}{\lambda} 
\frac{z_\alpha^\lambda}{\xi_\alpha(-\beta)^\lambda}
\bfa{b^T}_\alpha \bfa{q} = \frac{1}{\lambda} r^\lambda
 \left[\frac{\xi_\alpha(\theta)}{\xi_\alpha(-\beta)} 
\right]^\lambda \bfa{b^T}_\alpha \bfa{q}
\end{equation}

\nod where $\xi(\theta) = \cos(\theta) + p_\alpha \sin(\theta)$ and $\bfa{q}$
 is to be determined. The traction with respect to a radial plane at angle
 $\theta$ is given  by

\begin{equation}
\bfa{t} = r^{\lambda-1} \sum_{\alpha=1}^6
\left[\frac{\xi_\alpha(\theta)}{\xi_\alpha(-\beta)} \right]^\lambda
\bfa{b}_\alpha \bfa{b^T}_\alpha \bfa{q} = \frac{\lambda}{r} \phi
\end{equation}

\nod With the above form the traction condition is already
 satisfied on the bottom flank $\theta = -\beta$. Applying the condition on the
 top flank leads to a matrix equation

\begin{equation}
\bfa{K}(\lambda)\bfa{q}(\lambda) = 0
\end{equation}

\nod The appropriate value of $\lambda$ is determined by setting the
 determinant of the matrix equal to zero and solving the resulting equation
 numerically. In the range $0 < \lambda < 1$, two values can be found, 
corresponding to modes I and II, $\lambda^I$ and $\lambda^{II}$. For a given
 $\lambda$, the vector $\bfa{q}$
 is determined up to a normalization which is related to how one defines the 
stress intensity factor $K$. Thus we obtain expressions for the displacements 
which are used in the simulation to place the boundary atoms. In the crack
 case, because $\lambda^I$ and $\lambda^{II}$ are degenerate at the value 
$1/2$, the definition of modes I and II is a little subtle. The \{111\} 
plane is not a plane of symmetry of the cube, and thus one cannot expect to 
separate modes by their symmetry properties as in for example, the isotropic 
case; following Ref.~\onlinecite{Sih/Paris/Irwin:1965}, mode I is defined
as that for which $\sigma_{12}(\theta=0) = 0$ and mode II that for which 
$\sigma_{22}(\theta=0) = 0$.

For the purpose of the simulations described in this paper, we calculated the
 Stroh parameters as follows. For each potential, the elastic constants were
 determined by standard methods (straining the supercell, relaxing, measuring 
the relaxed energy per unit undeformed volume and fitting to a 
 parabola). This gives $C_{11}$, $C_{12}$, and $C_{44}$, which are the three
 independent constants for a cubic crystal. In the formulas for the
 displacements and stresses given above, the coordinate system is aligned with
 the notch (in that the negative $x$-axis bisects the notch itself) and not with
 the crystal axes. So we must transform the elastic constants accordingly. Once
 we have the transformed constants we can construct the Stroh matrices ${\bfa
 Q}$, ${\bfa R}$ and ${\bfa T}$, and
 compute the Stroh eigenvalues and eigenvectors as above.

\bibliography{silicon}

\end{document}